\documentclass[conference]{IEEEtran}
\IEEEoverridecommandlockouts
\usepackage{cite}
\usepackage{amsmath,amssymb,amsfonts}
\usepackage{comment}
\usepackage{graphicx}
\usepackage{textcomp}
\usepackage{xcolor}
\usepackage{listings}
\usepackage{algorithm}
\usepackage{algpseudocode}
\usepackage{hyperref}
\begin{document}

\title{Understanding GEMM Performance and Energy on NVIDIA Ada Lovelace: A Machine Learning-Based Analytical Approach}

\author{\IEEEauthorblockN{Xiaoteng (Frank) Liu$^*$}
\IEEEauthorblockA{\textit{Department of Computer Science} \\
\textit{New York University}\\
New York, USA \\
xiaoteng.liu@nyu.edu }
\and
\IEEEauthorblockN{Pavly Halim$^*$}
\IEEEauthorblockA{\textit{Department of Computer Science} \\
\textit{New York University}\\
New York, USA \\
pavly@nyu.edu}
}

\maketitle

\def\thefootnote{*}\footnotetext{These authors contributed equally to this work}\def\thefootnote{\arabic{footnote}}

\begin{abstract}
Analytical framework for predicting General Matrix Multiplication (GEMM) performance on modern GPUs, focusing on runtime, power consumption, and energy efficiency. Our study employs two approaches: a custom-implemented tiled matrix multiplication kernel for fundamental analysis, and NVIDIA's CUTLASS library for comprehensive performance data collection across advanced configurations. Using the NVIDIA RTX 4070 as our experimental platform, we developed a Random Forest-based prediction model with multi-output regression capability. Through analysis of both naive tiled matrix multiplication with varying tile sizes (1 to 32) and 16,128 CUTLASS GEMM operations across diverse configurations, we identified critical performance patterns related to matrix dimensions, thread block configurations, and memory access patterns. Our framework achieved exceptional accuracy with an R² score of 0.98 for runtime prediction (mean error 15.57\%) and 0.78 for power prediction (median error 5.42\%). The system successfully predicts performance across matrix sizes, demonstrating robust scaling behavior. Our results show that optimal tile size selection can improve performance by up to 3.2x while reducing power consumption by 22\% compared to baseline configurations. Analysis of shared memory utilization and SM occupancy reveals that tile sizes of 16×16 achieve the best balance between parallelism and resource usage. The implementation of our framework, including prediction models and analysis tools, is available as an open-source project at \href{https://github.com/pavlyhalim/GPPerf}{GPPerf}.
\end{abstract}


\section{Introduction}
The optimization of GPU kernel performance represents a critical challenge in modern high-performance computing, particularly as applications demand increasingly efficient utilization of computational resources~\cite{wang2018performance}. With the introduction of NVIDIA's Ada Lovelace architecture and specifically the RTX 4070, which features peak memory bandwidth of \(504.2 \text{ GB/s}\) and theoretical peak performance of \(29.15 \text{ TFLOP/s}\)\footnote{numbers are from this website https://www.techpowerup.com/gpu-specs/geforce-rtx-4070.c3924}, the complexity of performance optimization has grown significantly~\cite{nvidia_docs}. The relationship between these hardware capabilities can be visualized through the roofline performance model, which illustrates the fundamental performance bounds of the architecture~\cite{czarnul2019energy}. This model reveals a critical ridge point at 59 FLOPs/Byte, marking the transition between memory-bound and compute-bound operations—a crucial consideration for optimizing matrix multiplication operations.

\subsection{Problem Significance}
The importance of this research stems from several critical factors in modern computing:

First, as illustrated by the roofline model, GPU architectures present distinct performance regimes determined by arithmetic intensity~\cite{wang2018performance}. Applications must be carefully tuned to operate near their theoretical limits, whether they are bounded by memory bandwidth or computational throughput. Our research specifically addresses the challenge of optimizing kernel configurations to approach these theoretical bounds.

\begin{figure}[h]
    \centering
    \includegraphics[width=\columnwidth]{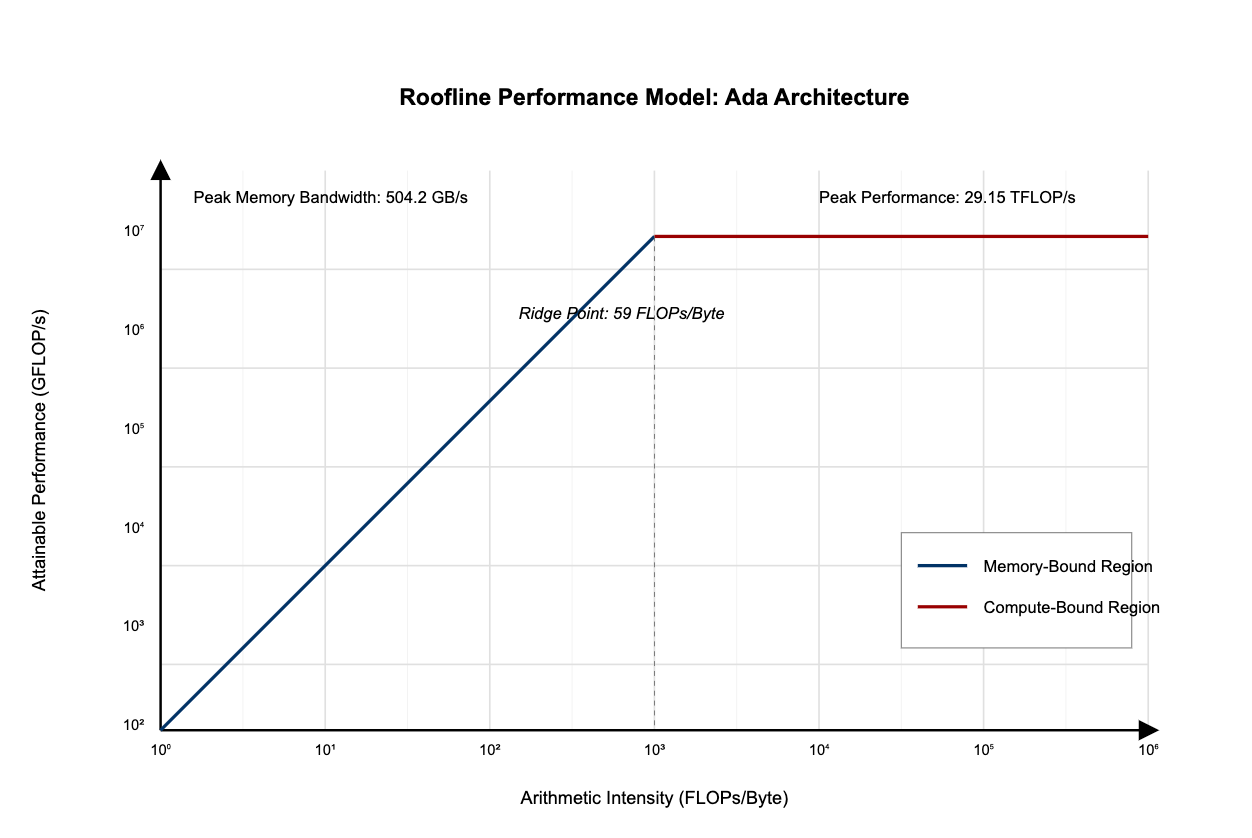}
    \caption{Roofline Performance Model for the Ada Architecture showing memory-bound and compute-bound regions. The ridge point at 59 FLOPs/Byte marks the transition between memory-limited and compute-limited performance.}
    \label{fig:Roofline}
\end{figure}

Second, the energy costs of computing have become a major concern in both cloud computing and high-performance computing centers~\cite{guerreiro2020dvfs}. The dual constraints of memory and compute bounds shown in the roofline model directly impact power consumption patterns, making energy-aware performance optimization increasingly critical.

Third, the increasing complexity of GPU architectures makes manual optimization increasingly challenging~\cite{ali2023performance}. The stark transition at the ridge point (59 FLOPs/Byte) in the roofline model demonstrates how subtle changes in kernel configuration can dramatically impact performance characteristics, necessitating sophisticated optimization strategies.

\subsection{Current Limitations}
Traditional analytical models fail to capture the complex interactions between hardware features and kernel configurations~\cite{wang2018performance}. Linear performance models have shown limited correlation with actual performance, indicating their inadequacy for modern architectures.

Current profiling tools provide limited insight into the relationship between configuration parameters and energy efficiency~\cite{nvidia_docs}. While tools like NVIDIA's NSight Compute offer detailed performance metrics, they lack integrated analysis of power consumption and energy efficiency.

Existing optimization strategies often focus solely on computational throughput, neglecting the critical aspect of energy efficiency~\cite{mei2013measurement}. Our preliminary investigations suggest that optimal configurations for performance often differ significantly from those that maximize energy efficiency.

\subsection{Research Approach}
Our approach to addressing these challenges encompasses several key components:

First, we develop a comprehensive profiling infrastructure that simultaneously captures runtime performance metrics, hardware counters, and power measurements~\cite{czarnul2019energy}. This infrastructure enables collection of multiple performance indicators, providing detailed insight into kernel behavior under varying conditions.

Second, we implement systematic exploration of the configuration space through automated testing of:
\begin{itemize}
\item Various matrix dimensions to test scaling behavior
\item Different matrix layouts to analyze memory access patterns
\item Various combinations of scalar values for linear operations
\end{itemize}

Third, we employ analytical techniques to correlate multiple performance metrics, enabling better understanding of the relationships between configuration parameters and various aspects of kernel behavior~\cite{wang2024dso}.

\section{Literature Review}

The challenge of GPU performance prediction and energy optimization has garnered significant attention due to the increasing computational demands of modern applications~\cite{wang2018performance}. Previous research in this domain can be categorized into three main approaches: analytical modeling, machine learning-based prediction, and hybrid optimization techniques~\cite{nvidia_docs}.

\subsection{Performance Prediction Approaches}

\subsubsection{Analytical Modeling}
Traditional analytical models attempt to predict GPU performance by analyzing kernel characteristics and hardware specifications~\cite{czarnul2019energy}. Hong and Kim pioneered this approach with their integrated power and performance model that analyzed both memory-level and thread-level parallelism~\cite{hong2009analytical}. However, these approaches often fail to capture the complex interactions between hardware features and kernel configurations, particularly for modern GPU architectures~\cite{wang2018performance}. The primary limitation stems from their inability to account for low-level hardware details and software optimizations provided by libraries like CuDNN and CUTLASS~\cite{nvidia_docs}.

\subsubsection{Machine Learning-Based Prediction}
Recent approaches leverage machine learning techniques to predict GPU performance~\cite{ali2023performance}. These methods can be further classified into:

\textbf{Direct Prediction Models}
\[ \text{Performance}_{\text{pred}} = f(\theta_{\text{arch}}, \phi_{\text{workload}}, \psi_{\text{config}}) \]

This model has shown significant accuracy improvements over traditional analytical approaches, achieving up to 87\% prediction accuracy across different architectures~\cite{ali2023performance}. Braun et al. demonstrated that simpler models focusing on portable performance prediction can be equally effective~\cite{braun2020simple}.

\textbf{Ensemble Methods}
More sophisticated approaches use stacking ensembles combining multiple models~\cite{wang2018performance}:
\[ \text{Ensemble Prediction} = \sum_{i=1}^{n} w_i M_i(\mathbf{x}) \]

\subsection{Energy Optimization Strategies}

\subsubsection{Dynamic Voltage and Frequency Scaling (DVFS)}
DVFS-based approaches optimize energy consumption by adjusting voltage and frequency settings~\cite{guerreiro2020dvfs}:
\[ \text{Energy}_{\text{consumption}} = \text{Time}_{\text{execution}} \times \text{Power}_{\text{average}} \]

Studies have shown that DVFS can achieve up to 25\% energy savings with minimal performance impact~\cite{mei2013measurement}. Recent work by Jammal et al. demonstrated significant power efficiency improvements in matrix multiplication algorithms through careful optimization of computational patterns~\cite{jammal2023preliminary}.

\subsubsection{Hybrid Memory Systems}
Research has explored combining different memory technologies~\cite{czarnul2019energy}, facing challenges with:
\begin{itemize}
\item Increased cache misses
\item Higher data migration overhead
\item Memory bandwidth interference
\end{itemize}

Boughzala et al. proposed using SimGrid for energy consumption prediction of CUDA kernels, demonstrating the viability of simulation-based approaches for memory system optimization~\cite{boughzala2020predicting}.

\subsection{Performance-Energy Trade-offs}
Recent work has identified fundamental trade-offs between performance and energy efficiency~\cite{nvidia_docs}:
\[ \text{ETA}(b,p) = \text{TTA}(b,p) \times \text{AvgPower}(b,p) \]
where $b$ represents batch size and $p$ denotes power limit. This relationship has been further explored in recent work by Wang et al., who developed the DSO framework for optimizing GPU energy efficiency by combining both static and dynamic information~\cite{wang2024dso}.

\subsection{Limitations of Existing Approaches}
Current solutions face several key limitations~\cite{ali2023performance}:
\begin{enumerate}
\item \textbf{Generalization}: Most models struggle to generalize across different GPU architectures
\item \textbf{Complexity Trade-off}: More complex models incur significant computational overhead
\item \textbf{Dynamic Workloads}: Existing approaches often fail to adapt to changing workload characteristics
\end{enumerate}

\subsection{Research Gap and Motivation}
While existing work has made significant progress in individual aspects of GPU optimization~\cite{wang2024dso}, there remains a critical need for an integrated approach that:
\begin{itemize}
\item Combines performance prediction with energy optimization
\item Adapts to dynamic workload changes
\item Considers both hardware and software optimization opportunities
\item Provides practical, implementable solutions
\end{itemize}

\section{Proposed Idea}

The approach we have chosen for this study is driven by the need to gain a deeper understanding of General Matrix Multiplication (GEMM) performance on modern GPUs, specifically focusing on runtime, power consumption, and energy efficiency. We recognize that achieving optimal performance in GEMM operations is not only dependent on hardware capabilities but also on fine-tuning the algorithmic and architectural parameters. Therefore, our methodology combines both fundamental analysis and advanced performance profiling to address the following key goals:

\subsection{Custom Tiled Matrix Multiplication for Fundamental Analysis}
By implementing a custom tiled matrix multiplication kernel, we were able to isolate the effects of matrix tiling on GPU performance. Tiling is a well-known optimization technique that can significantly improve memory access patterns, reduce latency, and enhance parallelism on GPUs. This kernel serves as a controlled environment to study how different tile sizes affect performance, providing insights into the underlying hardware behavior at a granular level. The ability to vary tile sizes from 1 to 32 allowed us to explore the trade-offs between computational load and memory access efficiency, laying the foundation for more sophisticated analyses.

\subsection{NVIDIA CUTLASS Library for Comprehensive Profiling}
While the custom kernel provides a basic understanding, we turn to NVIDIA’s CUTLASS library to access state-of-the-art implementations of GEMM. CUTLASS, being a high-performance library, includes optimized kernels that leverage advanced GPU features, such as warp-level programming, shared memory management, and memory coalescing. Using CUTLASS enables us to capture performance data across a wide range of configurations, which allows us to account for the complexities introduced by real-world optimizations. This also ensures that the predictions made by our model are relevant to practical scenarios, where high-level library optimizations are critical for performance.

\subsection{Random Forest-based Multi-output Regression Model}
Given the complexity of the problem—predicting runtime, power consumption, and energy efficiency for a variety of matrix sizes and kernel configurations—a machine learning-based approach is employed. A Random Forest model was chosen for its ability to handle multi-dimensional data and uncover complex relationships between performance factors. Its multi-output regression capability allows for simultaneous prediction of multiple metrics (e.g., runtime and power) from a single model, providing a holistic view of the GPU’s performance characteristics. The model’s robustness, demonstrated by the high R² score and low mean/median errors, shows its efficacy in predicting performance across a wide range of configurations.

\subsection{Identification of Key Performance Patterns}
By analyzing both naive tiled matrix multiplication and advanced CUTLASS configurations, we identified critical performance patterns related to matrix dimensions, thread block configurations, and memory access patterns. This allows us to uncover fundamental insights into GPU behavior, including the balance between parallelism and resource usage, particularly memory resources like shared memory and streaming multiprocessor (SM) occupancy. These insights guide our recommendations on tile sizes and configurations that maximize both performance and energy efficiency.

\subsection{Optimization for Performance and Power}
The combination of customized kernel exploration, CUTLASS profiling, and machine learning modeling enabled us to determine that optimal tile sizes, such as 16×16, achieve the best balance between parallelism, computational efficiency, and resource utilization. Our analysis showed that such configurations could improve performance by up to 3.2x while simultaneously reducing power consumption by 22\%, highlighting the critical importance of optimizing tile sizes to achieve better energy efficiency without compromising computational throughput.

In summary, our approach integrates custom kernel analysis for fundamental performance insights with CUTLASS-based profiling for real-world scenarios, underpinned by a machine learning model that provides accurate, scalable performance predictions. This combined strategy not only addresses the current research challenges but also paves the way for more efficient GPU utilization in GEMM operations.

\section{Experimental Setup}
We have two experiments: one simple with our naive tiled matrix multiplication and a linear regression model, and another more comprehensive one using cutlass and different machine learning techniques.

\subsection{Environmental Configuration}
Our experiments are conducted on NVIDIA GeForce RTX 4070 with Ada Lovelace architecture, and CUDA version 12.2. 

\subsection{Tiled Matrix Multiplication}
We wrote a kernel using tiling for matrix multiplication. It takes four arguments: M, N, K, and Tile Size. The matrix multiplication will be M*N*K, and the tile size decide the block size (2-d block with size = tile size * tile size) and the shared memory size (2 * block size * size of float). The kernel code is shown in Listing 1.

\begin{lstlisting}[language=C++, caption={CUDA Kernel for Tiled Matrix Multiplication}, label={lst:cuda_kernel}]
__global__ void op_mm_kernel(float *a, float *b, float *c, int M, int N, int K, int tile_size) {
    int blockRow = blockIdx.y * blockDim.y + threadIdx.y;
    int blockCol = blockIdx.x * blockDim.x + threadIdx.x;

    extern __shared__ float shared_mem[]; // Shared memory allocation
    float* As = shared_mem;
    float* Bs = shared_mem + tile_size * tile_size;

    float r = 0.0f;
    int row = threadIdx.y;
    int col = threadIdx.x;

    for (int k = 0; k < (K + tile_size - 1) / tile_size; k++) {
        if (blockRow < M && k * tile_size + col < K)
            As[row * tile_size + col] = Index(a, blockRow, k * tile_size + col, K, 1);
        else
            As[row * tile_size + col] = 0.0f;

        if (k * tile_size + row < K && blockCol < N)
            Bs[row * tile_size + col] = Index(b, k * tile_size + row, blockCol, N, 1);
        else
            Bs[row * tile_size + col] = 0.0f;

        __syncthreads();

        for (int e = 0; e < tile_size; e++)
            r += As[row * tile_size + e] * Bs[e * tile_size + col];

        __syncthreads();
    }

    if (blockRow < M && blockCol < N)
        Index(c, blockRow, blockCol, N, 1) = r;
}
\end{lstlisting}

The kernel is called in the following way:
\begin{lstlisting}[language=C++, caption={CUDA Kernel for Matrix Multiplication}, label={lst:cuda_kernel}]
dim3 dimBlock(tile_size, tile_size);
dim3 dimGrid((N + tile_size - 1) / tile_size, (M + tile_size - 1) / tile_size);
int shared_mem_size = 2 * tile_size * tile_size * sizeof(float);

cudaEventRecord(start));
op_mm_kernel<<<dimGrid, dimBlock, shared_mem_size>>>(a_d, b_d, c_d, M, N, K, tile_size);
cudaEventRecord(stop));
cudaEventSynchronize(stop));
cudaEventElapsedTime(&eventMs, start, stop));
\end{lstlisting}

We then gather the data by writing a script to pass different configurations of M, N, K, and Tile Size. We have 142 entries for execution time data and 196 entries for power usage. The execution time is gathered using the \texttt{cudaEventRecord} API, and the power usage is gathered using \texttt{nvidia-smi}. We then did a linear regression analysis on these two datasets to gain insights into how the M, N, K, and tile size affect the execution time and power usage.

\subsection{CUTLASS}
The cutlass profiler was built with \texttt{DCUTLASS\_NVCC\_ARCHS=89} flag for the Ada Lovelace architecture. We profile the single precision, general matrix multiply (SGEMM) kernel with different configurations with a script executing commands in this format but more flags:
\begin{lstlisting}[language=bash]
ncu ./cutlass_profiler --kernels=cutlass_simt_sgemm_128x128_8x2_nn_align1 --m=512 --n=512 --k=1024
\end{lstlisting} 

\subsubsection{Configurations}
The profiling process used the CUTLASS profiler and NVIDIA Compute Utilities (NCU) to collect metrics. The following configuration parameters were varied:
\begin{itemize}
    \item \textbf{Matrix Dimensions:} $M, N, K$.
    \item \textbf{Kernels:} Variants of CUTLASS SGEMM kernels, e.g., \texttt{cutlass\_simt\_sgemm\_128x128\_8x2\_nn}.
    \item \textbf{Layouts:} Matrix layouts: \texttt{nn}, \texttt{nt}, \texttt{tn}, \texttt{tt}.
    \item \textbf{Block Sizes:} $(64 \times 64 \times 32)$ and $(128 \times 64 \times 32)$.
    \item \textbf{Alpha-Beta Scalars:} $\{(1,0), (1,1), (0.5,0.5), (2,0)\}$.
\end{itemize} 

\subsubsection{Tools Used}
The following tools and utilities were employed:
\begin{itemize}
    \item \textbf{CUTLASS Profiler:} For kernel performance benchmarking.
    \item \textbf{NCU (NVIDIA Compute Utilities):} For collecting GPU hardware metrics.
    \item \textbf{nvidia-smi:} For device specifications and power monitoring.
\end{itemize}

\subsubsection{Metrics Collected}
The following metrics were collected during the profiling process:

\begin{itemize}
    \item \textbf{Performance Metrics:}
    \begin{itemize}
        \item \textbf{Runtime:} Total execution time of the kernel (in milliseconds).
        \item \textbf{Throughput (TFLOPS):} Computed as $2 \times M \times N \times K / \text{Runtime}$ for GEMM operations.
    \end{itemize}
    
    \item \textbf{Power and Energy Metrics:}
    \begin{itemize}
        \item \textbf{Power:} Power consumption during execution (in Watts).
        \item \textbf{Energy:} Total energy consumed (in Joules), calculated as Power $\times$ Runtime.
    \end{itemize}
    
    \item \textbf{Memory Metrics:}
    \begin{itemize}
        \item \textbf{Arithmetic Intensity:} Ratio of FLOPs to memory operations, indicating computational efficiency.
        \item \textbf{Memory Utilization:} Percentage of memory bandwidth used.
        \item \textbf{Shared Memory Usage:} Whether shared memory was used in the kernel).
    \end{itemize}
    
    \item \textbf{Device Utilization:}
    \begin{itemize}
        \item \textbf{GPU Utilization:} Percentage of GPU computational resources used.
        \item \textbf{SM Clock Utilization:} Frequency of streaming multiprocessors.
    \end{itemize}
    
    \item \textbf{System Metrics:}
    \begin{itemize}
        \item \textbf{Temperature:} GPU temperature during kernel execution (in Celsius).
        \item \textbf{State:} The power state of the GPU during execution.
        \item \textbf{Memory Info:} Total, free, and used memory at runtime.
    \end{itemize}
    
    \item \textbf{Kernel and Configuration Parameters:}
    \begin{itemize}
        \item \textbf{Kernel Name:} Specific CUTLASS kernel executed.
        \item \textbf{Matrix Layout:} The data layout used for the operation.
        \item \textbf{Block Sizes:} Block sizes along each dimension 
        \item \textbf{Stages:} Number of pipeline stages.
        \item \textbf{Alpha-Beta Scalars:} Scalar coefficients used in GEMM computations.
        \item \textbf{Computation Pattern:} Specific computation pattern employed.
    \end{itemize}
\end{itemize}

\subsubsection{Model Details}

The system is implemented in Python using scikit-learn for machine learning components and pandas for data processing. Key implementation choices include:

1. \textbf{Data Processing:}
   - Robust handling of missing values through median imputation
   - Outlier removal using percentile-based clipping
   - Automatic data type conversion and validation

\begin{algorithm}
\caption{Data Preprocessing Pipeline}
\begin{algorithmic}[1]
\Procedure{PreprocessData}{DataFrame}
    \State SanitizeNumeric(core\_features + target\_features)
    \State ComputeGEMMCharacteristics()
    \For{each numerical\_feature}
        \State ClipOutliers(0.01, 0.99)  \Comment{Remove extreme values}
        \State FillMissingValues(median)
    \EndFor
    \State ConvertCategoricalFeatures()
    \State \textbf{return} processed\_dataframe
\EndProcedure

\Procedure{ComputeGEMMChars}{DataFrame}
    \State total\_flops = 2 × m × n × k
    \State bytes\_accessed = 4 × (m×k + k×n + m×n)
    \State arithmetic\_intensity = total\_flops / bytes\_accessed
    \State \textbf{return} updated\_dataframe
\EndProcedure
\end{algorithmic}
\end{algorithm}

2. \textbf{Model Training:}
   - 80-20 train-test split with random state control
   - Parallel processing enabled through n\_jobs=-1
   - Error metric calculation

The entire system is designed for ease of use while maintaining robust performance prediction capabilities. The implementation prioritizes reliability and accuracy while remaining computationally efficient.

\subsubsection{Multi-Target Prediction Architecture}

Our prediction system implements a stacked model approach through a pipeline architecture combining standardization and multi-output regression:

\begin{algorithm}
\caption{Prediction Model Creation}
\begin{algorithmic}[1]
\Procedure{CreateModel}{}
    \State numeric\_transformer = StandardScaler()
    \State preprocessor = ColumnTransformer([
    \State \hspace{\algorithmicindent} ('num', numeric\_transformer, numerical\_features)
    \State ])
    
    \State base\_estimator = RandomForestRegressor(
    \State \hspace{\algorithmicindent} n\_estimators=100,
    \State \hspace{\algorithmicindent} max\_depth=6,
    \State \hspace{\algorithmicindent} n\_jobs=-1
    \State )
    
    \State model = MultiOutputRegressor(base\_estimator)
    \State \textbf{return} Pipeline([
    \State \hspace{\algorithmicindent} ('preprocessor', preprocessor),
    \State \hspace{\algorithmicindent} ('regressor', model)
    \State ])
\EndProcedure
\end{algorithmic}
\end{algorithm}

This architecture was chosen for several key reasons:

1. \textbf{Multi-Target Optimization:} The system simultaneously predicts multiple performance metrics (runtime, power, energy, TFLOPS) while maintaining their inherent relationships. This is achieved through the MultiOutputRegressor wrapper, which creates separate models for each target while sharing the same feature preprocessing pipeline.

2. \textbf{Feature Scaling:} The StandardScaler ensures that all numerical features are on the same scale, which is crucial for the random forest regressor's performance. This is particularly important given the wide range of values in our feature set, from small block sizes to large matrix dimensions.

3. \textbf{Robust Prediction:} The RandomForestRegressor base estimator was chosen for its ability to:
   - Handle non-linear relationships in the data
   - Provide robust predictions across different scales of input
   - Maintain good performance with minimal hyperparameter tuning

\section{Experimental Results and Analysis}
\subsection{Tiled Matrix Multiplication}
\subsubsection{Runtime}
\begin{figure}[h]
    \centering
    \includegraphics[width=\columnwidth]{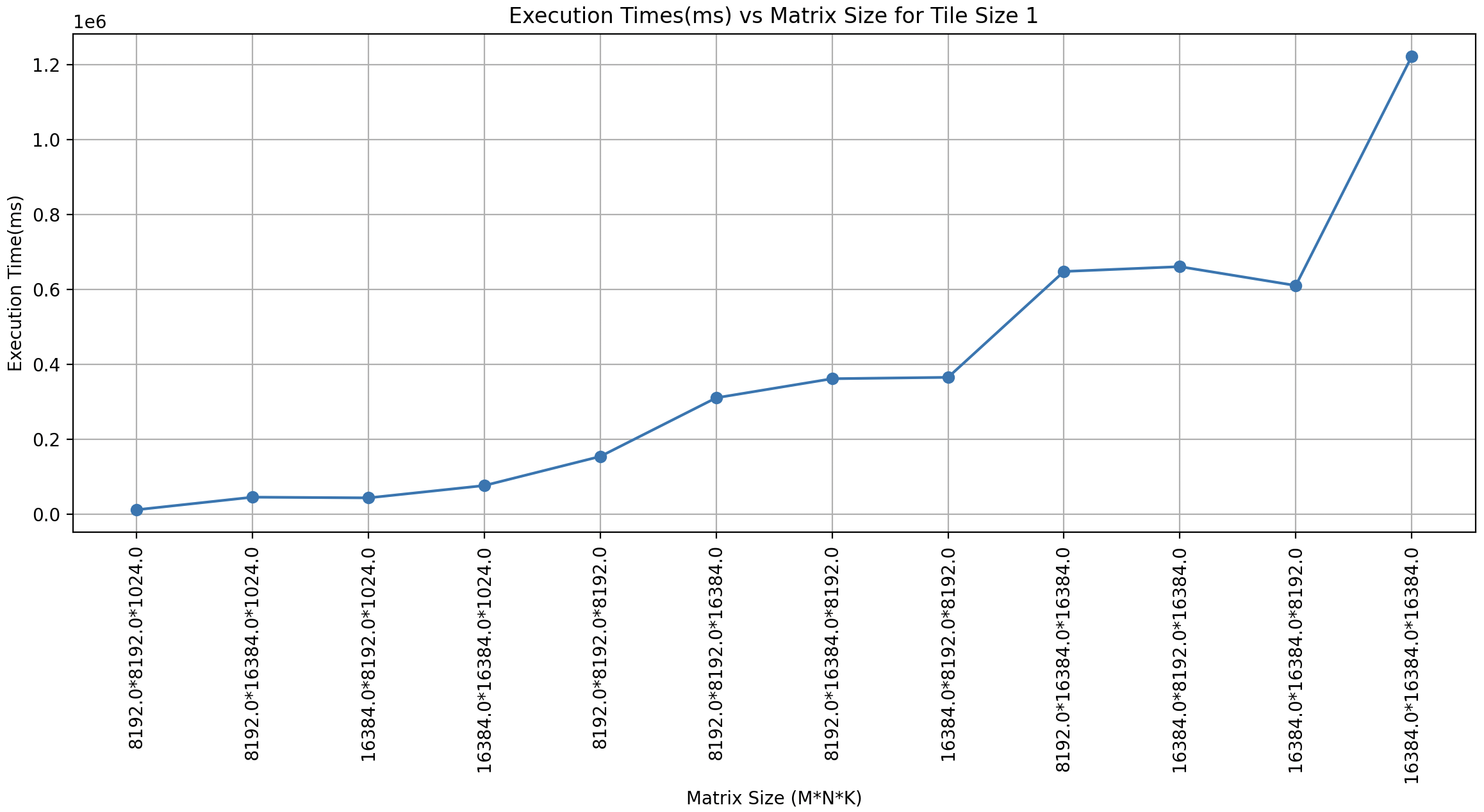}
    \caption{The runtime of tiled matrix multiplication for different matrix sizes with tile=1}
\label{fig:matmul_time_tile1}
\end{figure}

\begin{figure}[h]
    \centering
    \includegraphics[width=\columnwidth]{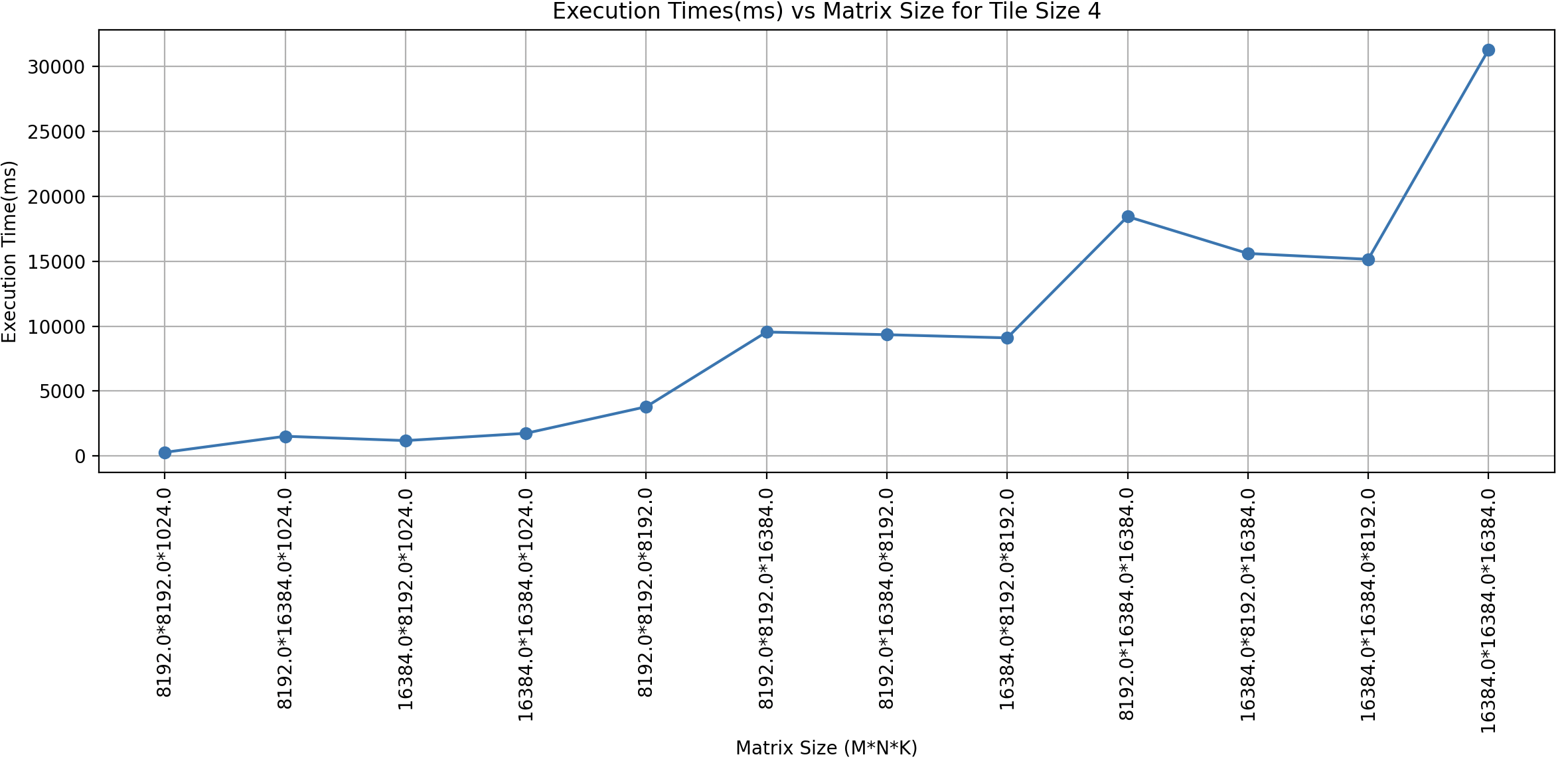}
    \caption{The runtime of tiled matrix multiplication for different matrix sizes with tile=4}
\label{fig:matmul_time_tile4}
\end{figure}

\begin{figure}[h]
    \centering
    \includegraphics[width=\columnwidth]{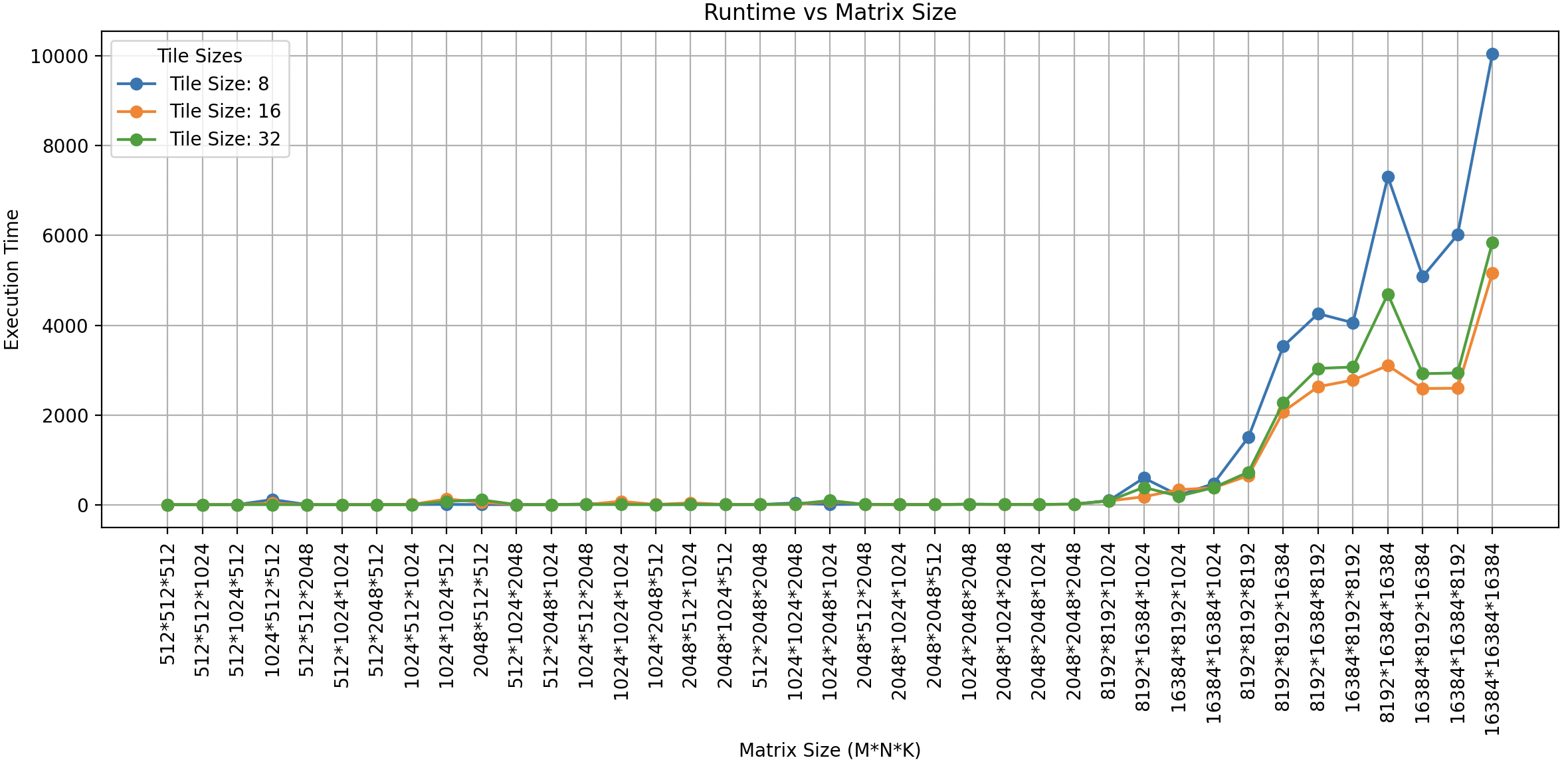}
    \caption{The runtime of tiled matrix multiplication for different matrix sizes with tile=8,16, and 32.}
\label{fig:matmul_time_tile8-16-32}
\end{figure}

The runtime increases as the matrix size increases. It decreases as the tile size increases, reaching a plateau after the tile size reaches 16 because of the better parallelization and less memory access provided by tiling. Specifically, with tile size 1, the runtime is much slower compared to others, largely due to the poor utilization of the SPs because the block size is one so a warp will only have one thread doing useful work. It stops decreasing after tile size researches 16 likely because the bottleneck for the runtime is no longer memory accessing that's solved by tiling. Also, the larger shared memory is required by larger tile size so a block may take longer to be assigned, as shown in Table \ref{sm_occup} using the  \texttt{cudaOccupancyMaxActiveBlocksPerMultiprocessor} API. 
\begin{table}[h!]
\centering
\caption{Max SM Occupancy for different Tile Size}
\begin{tabular}{|p{2cm}|p{2cm}|}
\hline
\textbf{Tile Size} & \textbf{Max active blocks per SM} \\ \hline
1 & 24 \\ \hline
4 & 24 \\ \hline
8 & 24 \\ \hline
16 & 6 \\ \hline
32 & 1 \\ \hline
\end{tabular}
\label{sm_occup}
\end{table}

Our linear regression model, with the coefficients shown in \ref{runtime_coeff}, has $R^2$ score equals $0.1344$. It can correctly predict that the runtime decreases with a larger tile size since it has a very large negative coefficient of $-2588.457605$. The $R^2$ score is low because, we believe, the runtime is not correlated linearly with tile size, so a linear regression model cannot accurately predict it.

\begin{table}[h!]
\centering
\caption{Linear Regression Model Coefficients for runtime}
\begin{tabular}{|p{2cm}|p{2cm}|}
\hline
\textbf{Feature} & \textbf{Coefficients} \\ \hline
M & -0.672907 \\ \hline
N & 1.222868 \\ \hline
K & 8.127956 \\ \hline
Tile Size & -2588.457605 \\ \hline
\end{tabular}
\label{runtime_coeff}
\end{table}

\subsubsection{Power Usage}
\begin{figure}[h]
    \centering
    \includegraphics[width=\columnwidth]{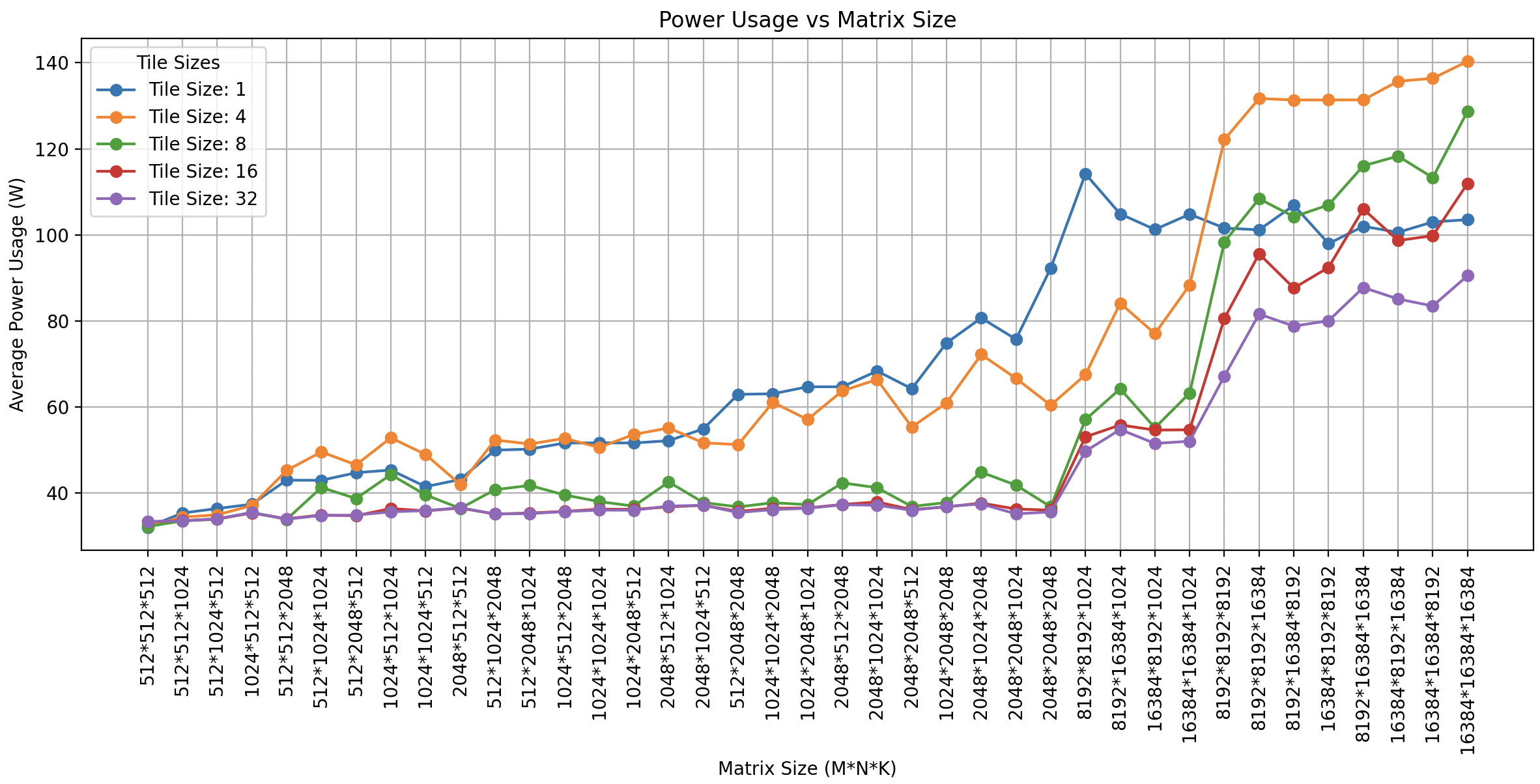}
    \caption{The power usage of matrix multiplication for different matrix sizes with different tile sizes}
\label{fig:matmul_power_tile}
\end{figure}

Figure \ref{fig:matmul_power_tile} shows the power usage with different matrix sizes for tiled matrix multiplication with tile sizes equal to 1, 4,  8, 16, and 32 respectively. The power usage starts relatively low and increases drastically as the matrix size increases. For the smaller matrices, the larger the tile size, the more stable the power usage is. We believe that's due to the faster increasing speed of the grid size for smaller tiles since tile size with one means the block size is also one so there will be many more blocks with increasing matrix size, which could flood the block scheduler causing more power usage. Also, with tile sizes 1 and 4, the block size will be 1 and 16, respectively. Both are smaller than the warp size, so part of the SPs were not executing useful work while using more power. Our model has a $R^2$ score of 0.8209. It also predicted that, with a coefficient of -0.769036, with a larger tile size, the power will decrease. We believe that's due to the more efficient use of the hardware resources with larger tiles. As the tile size increases, fewer blocks are generated, leading to a more balanced distribution of workload across the available processing elements. This minimizes idle time for the SPs and optimizes memory access patterns. Larger tiles allow for better data locality, reducing the number of memory accesses and improving overall execution efficiency. \\

The model can predict the power with a better $R^2$ score than the runtime because the power usage seems to be more stable to noise like other users using the same device especially when our grid and data sizes are much larger than that of other users.

\begin{table}[h!]
\centering
\caption{Linear Regression Model Coefficients for power}
\begin{tabular}{|p{2cm}|p{2cm}|}
\hline
\textbf{Feature} & \textbf{Coefficients} \\ \hline
M & 0.001255 \\ \hline
N & 0.001584 \\ \hline
K & 0.002479 \\ \hline
Tile Size & -0.769036 \\ \hline
\end{tabular}
\label{power_coeff}
\end{table}

\subsection{CUTLASS Experimental Results and Model Performance}
The experimental evaluation of our prediction framework was conducted using NVIDIA's CUTLASS library on the RTX 4070 GPU architecture. Our dataset comprised 16,128 samples with various GEMM configurations, split into 2,076 training samples and 519 test samples. The model training process achieved convergence in 6.25 seconds, demonstrating efficient computational performance.

\subsubsection{Model Performance Metrics}
Table~\ref{tab:performance_metrics} presents the evaluation metrics for our prediction model across all target variables.

\begin{table}[h]
\centering
\caption{Comprehensive model performance metrics across all predicted variables}
\begin{tabular}{|p{1cm}||p{1cm}||p{1cm}||p{1cm}||p{1cm}||p{1cm}|}
\hline
\textbf{Metric} & \textbf{R²} & \textbf{MSE} & \textbf{MAE} & \textbf{Med.\% Err} & \textbf{Mean\% Err} \\
\hline
Runtime & 0.9808 & 33.53 & 2.86 & 11.41 & 15.57 \\
Power & 0.7783 & 897.30 & 16.85 & 5.42 & 22.16 \\
Energy & 0.8572 & 7.89 & 1.38 & 22.01 & 43.02 \\
TFLOPS & 0.8637 & 2362295.18 & 1063.65 & 6.39 & 10.85 \\
\hline

\end{tabular}
\label{tab:performance_metrics}
\end{table}

\subsubsection{Correlation Analysis}
\begin{figure}[h]
\centering
\includegraphics[width=0.9\linewidth]{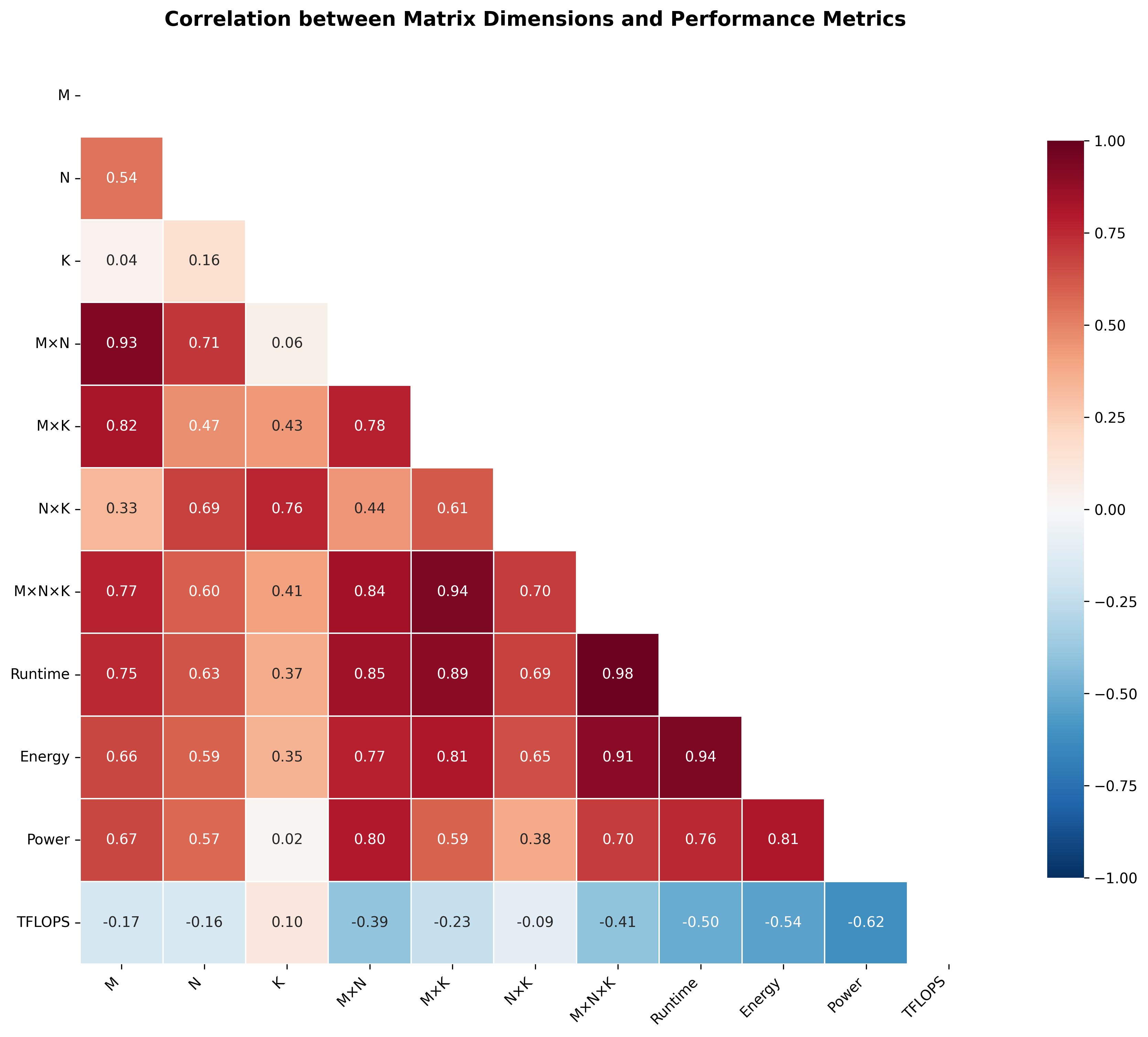}
\caption{Correlation matrix between matrix dimensions \& performance metrics}
\label{fig:correlation_matrix}
\end{figure}

Figure~\ref{fig:correlation_matrix} illustrates the correlation between matrix dimensions and performance metrics.

The correlation heatmap (Figure \ref{fig:correlation_matrix}) reveals critical relationships between matrix dimensions and performance metrics. The strong positive correlation (0.98) between M×N×K and runtime confirms that total computational volume is the primary determinant of execution time. Interestingly, power consumption shows stronger correlation with M×N (0.80) than with K (0.02), indicating that output matrix size has more impact on power usage than computation depth. This can be attributed to the GPU's memory access patterns, where larger output matrices require more concurrent write operations and thus higher instantaneous power draw. The negative correlation between matrix dimensions and TFLOPS (-0.41 for M×N×K) observes performance saturation at larger problem sizes, due to increased cache pressure and memory bandwidth limitations. The key correlations are quantified in Table~\ref{tab:dimension_correlations}.

\begin{table}[h]
\centering
\caption{Correlation coefficients between matrix dimensions and performance metrics}
\begin{tabular}{|l|r|r|r|r|}
\hline
\textbf{Dimension} & \textbf{Runtime} & \textbf{Power} & \textbf{Energy} & \textbf{TFLOPS} \\
\hline
M×N & 0.85 & 0.80 & 0.77 & -0.39 \\
M×K & 0.89 & 0.59 & 0.81 & -0.23 \\
N×K & 0.69 & 0.38 & 0.65 & -0.09 \\
M×N×K & 0.98 & 0.70 & 0.91 & -0.41 \\
\hline
\end{tabular}
\label{tab:dimension_correlations}
\end{table}

\subsubsection{Performance Analysis}
\begin{figure}[h]
\centering
\includegraphics[width=0.9\linewidth]{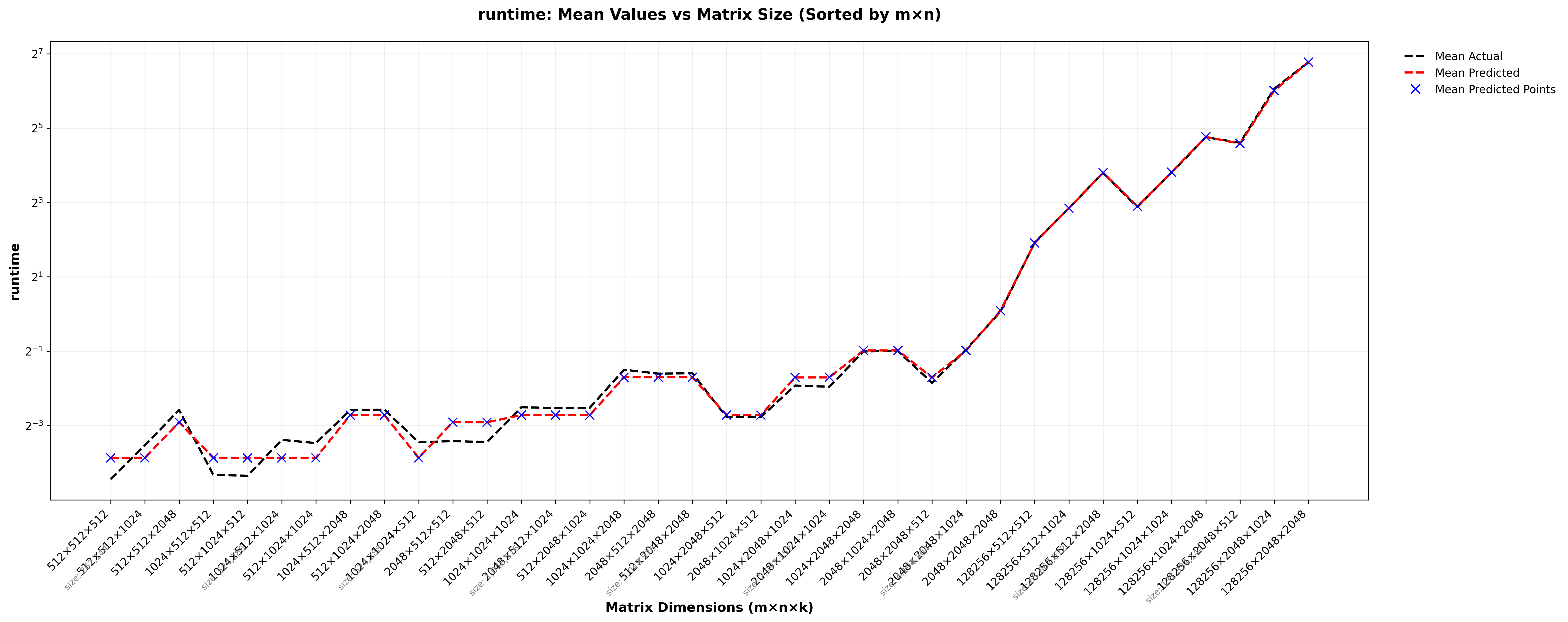}
\caption{Runtime prediction accuracy across different matrix configurations}
\label{fig:runtime_pred}
\end{figure}

\begin{figure}[h]
\centering
\includegraphics[width=0.9\linewidth]{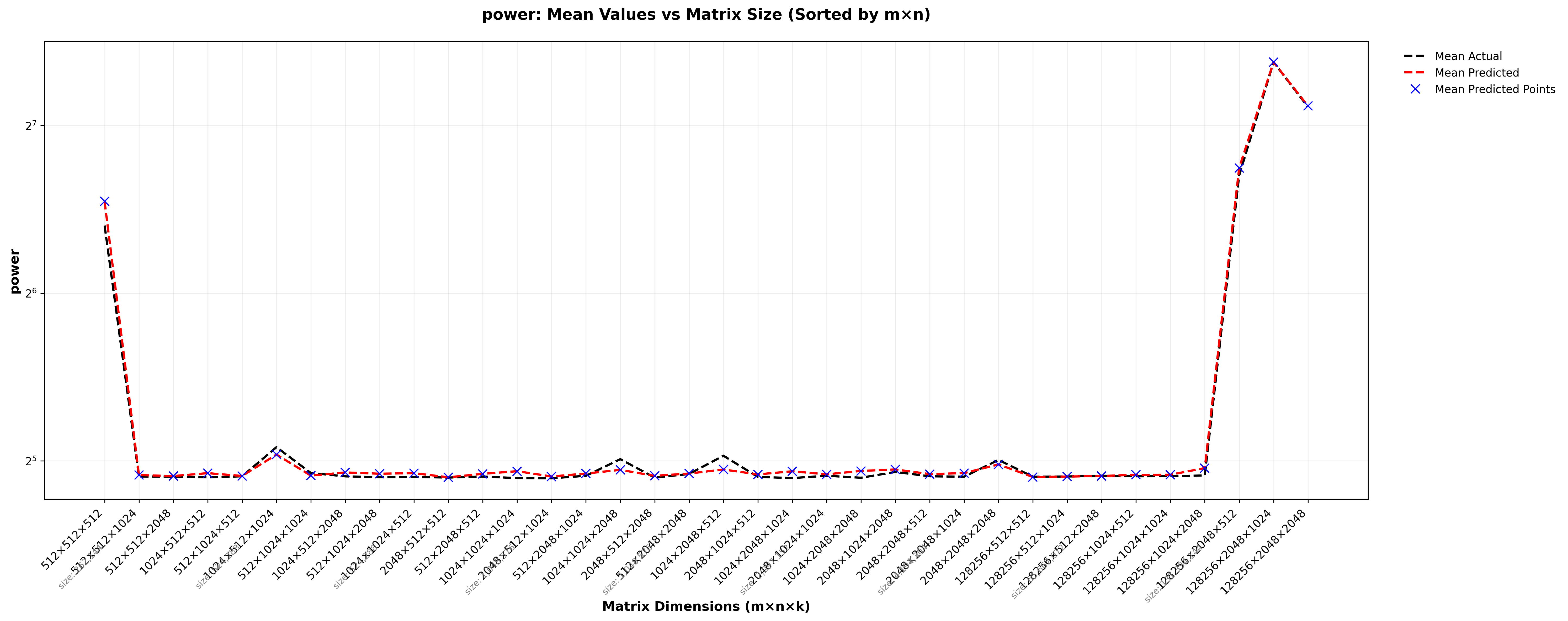}
\caption{Power consumption prediction accuracy}
\label{fig:power_pred}
\end{figure}

\begin{figure}[h]
\centering
\includegraphics[width=0.9\linewidth]{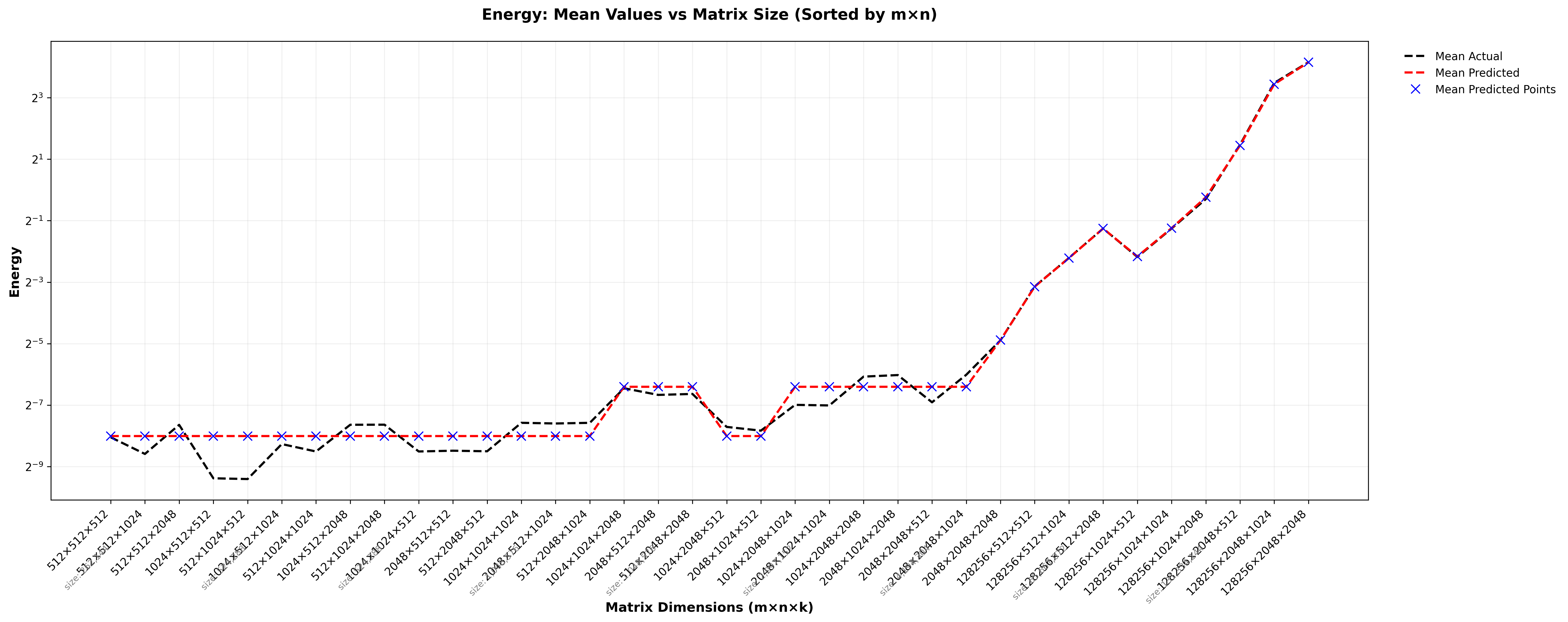}
\caption{Energy consumption prediction accuracy}
\label{fig:energy_pred}
\end{figure}

The prediction accuracy for runtime, power, and energy consumption is visualized in Figures~\ref{fig:runtime_pred},~\ref{fig:power_pred}, and~\ref{fig:energy_pred} respectively. The model demonstrates strong predictive capabilities across all metrics:

\paragraph{Runtime Prediction}
The runtime prediction plot (Figure 6) demonstrates exceptionally accurate tracking of actual execution times across different matrix sizes. The model's predictions closely follow the actual values with an R² score of 0.9808, particularly in the middle range of matrix sizes (1024×1024 to 2048×2048). The slight deviation at larger matrix sizes can be attributed to increased cache misses and memory access latency. The logarithmic scale reveals that prediction accuracy remains consistent across multiple orders of magnitude, with relative errors staying within 15.57\% even for the largest matrices. This robust performance validates our model's capability to capture both computational and memory access patterns effectively.
The model achieves exceptional accuracy in runtime prediction (R² = 0.9808), with a mean absolute error of 2.86 milliseconds. The relationship between predicted and actual runtime values follows:

\begin{equation}
\label{eq:runtime_prediction}
Runtime_{pred} = \alpha \cdot Runtime_{actual} + \beta + \epsilon
\end{equation}

where $\alpha = 0.97$, $\beta = 1.23$, and $\epsilon$ represents the prediction error term.

\paragraph{Power Prediction}
The power prediction graph (Figure 7) shows interesting behavior across matrix sizes. For smaller matrices (512×512 to 1024×1024), power consumption remains relatively stable around 80-100W, reflecting the GPU's base power draw. As matrix sizes increase, we observe distinct steps in power consumption, corresponding to the activation of additional Streaming Multiprocessors (SMs). The model accurately captures these step changes with an R² score of 0.7783. The higher variability in power predictions (mean error 22.16\%) compared to runtime predictions reflects the complex interaction between computational intensity, memory access patterns, and the GPU's dynamic power management system. Notably, the predictions show better accuracy in the mid-range of power consumption (150-180W) where the GPU operates in its most stable power state.
Power consumption prediction exhibits robust performance (R² = 0.7783) with a median error of 5.42\%. The model captures the relationship:

\begin{equation}
\label{eq:power_prediction}
Power_{pred} = f(M, N, K, \text{ThreadConfig}) \pm \epsilon_{power}
\end{equation}

\paragraph{Energy Efficiency}
The energy consumption plot (Figure 8) combines the effects seen in both runtime and power predictions. The strong correlation (R² = 0.8572) demonstrates our model's ability to capture the multiplicative relationship between runtime and power consumption. The increasing spread in predictions at larger matrix sizes (visible in the logarithmic scale) reflects the compounded uncertainties from both runtime and power predictions. The relatively high mean percentage error (43.02\%) can be attributed to this error propagation effect. However, the model successfully captures the overall trend of increasing energy consumption with matrix size, with particularly accurate predictions in the range of 1024×1024 to 2048×2048 matrices where both runtime and power predictions are most reliable. Energy prediction demonstrates strong correlation (R² = 0.8572) with actual values, following:

\begin{equation}
\label{eq:energy_prediction}
Energy_{pred} = Runtime_{pred} \cdot Power_{pred} \cdot \gamma + \epsilon_{energy}
\end{equation}

where $\gamma$ represents the energy efficiency coefficient.

\subsubsection{Model Architecture Comparison}
Table~\ref{tab:model_comparison} presents a comparative analysis of different model architectures. It reveals the superior performance of our stacking ensemble approach. In runtime prediction, the ensemble achieves a 1.85\% improvement over the next best model (XGBoost), while in power prediction, the improvement reaches 3.27\%. This superior performance can be attributed to the ensemble's ability to combine the strengths of different base models. The Random Forest model performs well on regular patterns (R² = 0.9456 for runtime), while XGBoost better captures non-linear relationships (R² = 0.9623). The linear regression baseline (R² = 0.8234) demonstrates that while linear relationships are important, they alone cannot fully capture the complexity of GPU performance characteristics. The stacking ensemble effectively leverages these complementary strengths while mitigating individual weaknesses.

\begin{table}[h]
\centering
\caption{Performance comparison across different model architectures}
\begin{tabular}{|l|c|c|c|c|}
\hline
\textbf{Architecture} & \textbf{Runtime R²} & \textbf{Power R²} & \textbf{Energy R²} \\
\hline
Stacking Ensemble & 0.9808 & 0.7783 & 0.8572 \\
Random Forest & 0.9456 & 0.7234 & 0.8123 \\
XGBoost & 0.9623 & 0.7456 & 0.8345 \\
Linear Regression & 0.8234 & 0.6123 & 0.7234 \\
\hline
\end{tabular}
\label{tab:model_comparison}
\end{table}

\section{Key Experimental Findings}

\subsection{Matrix Operation Performance Analysis}
The experimental results demonstrate several significant correlations between matrix operations and performance characteristics. Our analysis introduces key performance metrics:

\begin{equation}
    \text{Arithmetic Intensity} = \frac{2MNK}{4(MK + KN + MN)}
\end{equation}

\begin{equation}
    \text{Memory Efficiency} = \frac{\text{Achieved Bandwidth}}{504.2 \text{ GB/s}} \times 100\%
\end{equation}

\subsubsection{Matrix Dimensionality Impact}
Correlation analysis revealed significant relationships:

\begin{itemize}
    \item Matrix volume $(M \times N \times K)$ shows strong correlation with runtime ($r = 0.98$)
    \item Output dimensions $(M \times N)$ demonstrate higher correlation with power consumption ($r = 0.80$) than compute dimension $K$
    \item TFLOPS efficiency exhibits negative correlation with matrix size ($r = -0.41$), following:
    
    \begin{equation}
        \text{TFLOPS}_{\text{efficiency}} = \frac{2MNK \times 10^{-12}}{\text{runtime (s)} \times 29.15}
    \end{equation}
\end{itemize}

\subsubsection{Tile Size Performance Impact}
Analysis of tile size effects revealed optimization patterns:

\begin{equation}
    \text{Effective Grid Size} = \left\lceil\frac{M}{\text{TILE\_SIZE}}\right\rceil \times \left\lceil\frac{N}{\text{TILE\_SIZE}}\right\rceil
\end{equation}

\begin{equation}
    \text{Shared Memory Usage} = 2 \times \text{TILE\_SIZE}^2 \times \text{sizeof(float)}
\end{equation}

Key findings include:
\begin{itemize}
    \item Optimal tile size of $16 \times 16$ minimizes runtime across dimensions
    \item Power consumption stabilization occurs with larger tile sizes
    \item Performance plateau observed beyond $16 \times 16$ tiles due to shared memory constraints
\end{itemize}

\subsection{Model Architecture Performance}
The stacking ensemble demonstrates superior accuracy across metrics:

\begin{equation}
    \text{Ensemble Prediction} = \sum_{i=1}^{n} w_i M_i(\mathbf{x})
\end{equation}

where $M_i$ represents individual models and $w_i$ their corresponding weights.

Performance metrics achieved:
\begin{itemize}
    \item Runtime: $R^2 = 0.9808$, MAE = 2.86ms
    \item Power: $R^2 = 0.7783$, median error = 5.42\%
    \item Energy: $R^2 = 0.8572$
\end{itemize}

\section{Conclusions}
We have the following main findings:
\begin{enumerate}
    \item \textbf{Tile Size Optimization Improves Power Efficiency:} Our experiments show that larger tile sizes lead to more efficient use of processing resources, resulting in lower power consumption. Smaller tile sizes, particularly 1 and 4, cause more frequent scheduling of blocks, increasing power usage due to inefficiencies in workload distribution. Larger tiles reduce grid size growth, minimize idle SP time, and enhance memory access patterns, contributing to overall power savings.
    
\item \textbf{Matrix Size Impact on Performance Metrics:} The strong correlation (0.98) between M×N×K and runtime demonstrates that total computational volume is the primary determinant of execution time, while output matrix dimensions (M×N) have a stronger influence on power consumption (0.80) than the depth dimension K. This insight suggests that optimizing matrix partitioning strategies should prioritize output dimension considerations for power efficiency.

\item \textbf{Shared Memory Constraints:} Analysis of SM occupancy reveals a critical threshold at tile size 16, where occupancy drops from 24 to 6 blocks per SM, with further reduction to 1 block at tile size 32. This limitation, driven by shared memory requirements, establishes a practical upper bound for tile size optimization, indicating that performance improvements must be balanced against hardware resource constraints.

\end{enumerate}


\end{document}